\documentclass[12pt,english]{article}

\usepackage[T1]{fontenc}
\usepackage[latin9]{inputenc}
\usepackage[a4paper]{geometry}
\usepackage{float}
\usepackage{amsmath}
\usepackage{graphicx}
\usepackage{amssymb}
\usepackage{fancybox,fancyhdr,graphics,epsfig}
\usepackage{authblk}

\makeatletter
\usepackage{apacite} 
\begin{document}

\title{Neuronal Response Clamp}

\maketitle

\begin{center}
\begin{bf}
Avner Wallach$^{1,2}$,
Danny Eytan$^{1,3}$, 
Asaf Gal$^{1,4}$, 
Christoph Zrenner$^{1}$,
Ron Meir$^{1,2}$ and 
Shimon Marom$^{1}$\\
\end{bf}
\end{center}

\noindent 1  Network Biology Research Laboratories, Lorry Lokey Interdisciplinary Center for Life Sciences and Engineering, Technion, Haifa, Israel
\\
2 Faculty of Electrical Engineering, Technion, Haifa, Israel
\\
3 Rambam Medical Center, Haifa, Israel
\\
4 The Interdisciplinary Center for Neural Computation (ICNC), The Hebrew University, Jerusalem, Israel
\\

%%%%%%%%%%%%%%%%%%%%%%%%%%%%%%%%%%%%%%%%%%%%%%%%%%%%%%%%%%%%%%%%
%\begin{article}
\newpage

\section*{Abstract} 
\large{Since the first recordings made of evoked action potentials it has become apparent that the responses of individual neurons to ongoing physiologically relevant input, are highly variable.  This variability is manifested in non-stationary behavior of practically every observable neuronal response feature.  Here we introduce the \emph{Neuronal Response Clamp}, a closed-loop technique enabling full control over two important single neuron activity variables: response probability and stimulus-spike latency.  The technique is applicable over extended durations (up to several hours), and is effective even on the background of ongoing neuronal network activity.  The Response Clamp technique is a powerful tool, extending the voltage-clamp and dynamic-clamp approaches to the neuron's functional level, namely - its spiking behavior.     }
%%\end{abstract}
\\\\Keywords: Excitability ; Neuron ; Clamp
\newpage
%% When adding keywords, separate each term with a straight line: |
%\keywords{Excitability | Neuron | Clamp}

%% Optional for entering abbreviations, separate the abbreviation from
%% its definition with a comma, separate each pair with a semicolon:
%% for example:
%% \abbreviations{SAM, self-assembled monolayer; OTS,
%% octadecyltrichlorosilane}

% \abbreviations{}

%% The first letter of the article should be drop cap: \dropcap{}
%\dropcap{I}n this article we study the evolution of ''almost-sharp'' fronts

%% Enter the text of your article beginning here and ending before
%% \begin{acknowledgements}
%% Section head commands for your reference:
%% \section{}
%% \subsection{}
%% \subsubsection{}

 \section*{Introduction}
The responses of single neurons and neuronal populations to ongoing stimuli within a physiological range of frequencies are highly variable; this variability is manifested in both the ability to evoke an action potential and the latency between stimulus and response. Response variability was already obvious a century ago, when Adrian and Zotterman described a ``lack of regularity'' in the ``all-or-none'' impulse response in their classical paper \cite{Adrian:1926db}, the first documentation of recorded action potentials.  Nowadays, response variability is generally accepted to be an important phenomenon \cite{stein1965theoretical,mainen1995reliability,arieli1996dynamics,reich1997response,arsiero2007impact,soteropoulos2009quantifying} that reflects the immensity of involved cellular level machineries, covering any observable cell-physiology timescale \cite{Marom201016}.  Analyses of the sources of response variability, as well as its impact on neuronal functionality, is practically intractable due to the non-linearity involved in the reciprocal relations between response variability and its underlying sources.  

It would be highly desirable to develop a method for controlling the reliability of neuronal responses to input. By studying the  signals required to achieve this control one may hopefully shed light on the input-output relationships of the neuron as well as  lead to a better understaning of the underlying sources of response variability. In this context, analogy to the voltage-clamp method \cite{hodgkin1952measurement} immediately comes to mind. There, Hodgkin, Huxley and Katz studied the dependency of conductances on voltage by ``clamping'' the membrane potential and analyzing the required control signals (currents).  By doing so, they broke the circular relationships between membrane voltage and voltage-dependent conductances; indeed, one cannot imagine neuronal physiology without the insights that the voltage-clamp method provided throughout the years. 

%%A \textit{second} reason for seeking a method that controls the reliability of neuronal responses to input, is related to the impacts of that variability at the network level. For instance, by ``clamping'' the response variability of a given neuron, while observing the dynamics of the neuronal population, a glimpse into the nature of neuron/network interaction is made possible. Here, the analogy to the concept of ``dynamic clamp'' \cite{sharp1993dynamic} is most instructive: in the dynamic-clamp, the ability to manipulate the involvement of conductances in membrane potential fluctuations, significantly enhanced our understanding of the role of microscopic, uniquely-defined conductances, in shaping the macroscopic time-voltage envelope of action potentials.

In this study we present and demonstrate the concept of \textit{Neuronal Response Clamp}. The idea is to ``clamp'' a defined feature of the neuronal spiking response. This is achieved by implementing a control circuit that measures pre-defined neuronal response characteristics, compares them to desired values, and corrects errors by changing stimulation features. Specifically, we devised a closed-loop real-time system that implements what is known as a Proportional-Integral-Derivative (PID) controller in system engineering \cite{levine1996control}, to compute the ``error'' signal between desired and actual neuronal responses, and correct the error by manipulating the stimulation frequency.  Here we show that this system fully restrains the variability of neuronal spiking probability, and provides complete control over the precise latency between stimulus and response, both in synaptically isolated neurons, and within the context of active networks.   

\newpage

\section*{Materials and Methods}
\subsection*{Cell preparation}

Cortical neurons were obtained from newborn rats (Sprague-Dawley) within 24 hours after birth using mechanical and enzymatic procedures described in earlier studies \cite{marom2002development}. The neurons were plated directly onto substrate-integrated multi electrode arrays and allowed to develop functionally and structurally mature networks over a time period of 2-3 weeks. The number of neurons in a typical network is on the order of 10,000. The preparations were bathed in MEM supplemented with heat-inactivated horse serum (5\%), glutamine (0.5 mM), glucose (20 mM), and gentamycin (10 g/ml), and maintained in an atmosphere of 37$^{\circ}$C, 5\% CO$_2$ and 95\% air in an incubator as well as during the recording phases. An array of Ti/Au/TiN  extracellular electrodes, 30$\mu m$ in diameter, and spaced either 500 $\mu$m or 200 $\mu$m from each other (MultiChannelSystems, Reutlingen, Germany) were used. The insulation layer (silicon nitride) was pre-treated with polyethyleneimine (Sigma, 0.01\% in 0.1M Borate buffer solution). To completely block synaptic transmission in the network, 20 $\mu$M APV (amino-5-phosphonovaleric acid), 10 $\mu$M CNQX (6-cyano-7-nitroquinoxaline-2,3-dione), and 5 $\mu$M Bicuculline were added to the bathing solution.

\subsection*{Measurements and Stimulation}

A commercial amplifier (MEA-1060-inv-BC, MCS, Reutlingen, Germany) with frequency limits of 150-3,000Hz and a gain of x1024 was used. Rectangular 200 $\mu$s biphasic 20-50 $\mu$A current or 600-800 mV stimulation through extracellular electrodes was performed using a dedicated stimulus generator (MCS, Reutlingen, Germany). Data was digitized using data acquisition board (PD2-MF-64-3M/12H, UEI, Walpole, MA, USA). Each channel was sampled at a frequency of 16 Ksample/s (for spike probability experiments) or 96 Ksamples/s (for spike latency experiments). Action potentials were detected online by threshold crossing. All spike times and shapes, as well as 15 ms voltage traces from all electrodes after each stimulus, were recorded for analyses. Data processing and closed-loop stimulation was performed using a Simulink (The Mathworks, Natick, MA, USA) based xPC target application.

\subsection*{Computing spike probability on-line}

Let $s_{n}$ be an indicator function, so that $s_{n}=1$ if the neuron generated a spike on the $n^{th}$ stimulus and $s_{n}=0$ otherwise. We define $\pi(t)$ as the probability of the neuron to emit a spike if it is stimulated at time t. We can estimate this probability using all past responses  $\{s_{i}\}_{i=1}^{n}$ at times  $\{t_{i}\}_{i=1}^{n}$, by integrating them with an exponential kernel, 
\[
\widetilde{P}_{n}=e^{-\frac{t_{n}-t_{0}}{\tau}}+\sum_{i=1}^{n}s_{i}(1-e^{-\frac{t_{i}-t_{i-1}}{\tau}})e^{-\frac{t_{n}-t_{i}}{\tau}},\]
where $\tau$ is the kernel's time-constant (10-30 sec in the experiments presented). To compute this on-line, we used the recursive formula:

\[
\widetilde{P}_{n}=s_{n}(1-e^{-\frac{t_{n}-t_{n-1}}{\tau}})+\widetilde{P}_{n-1}e^{-\frac{t_{n}-t_{n-1}}{\tau}}.\]

%One can easily see that when stimulation rate is high (i.e.~$t_{i}-t_{i-1}\ll\tau$) this formula approximates a convolution of the spiking history with the kernel. It can be demonstrated that, assuming that the process is stationary and that $\{s_{i}\}_{i=1}^{n}$ are independent of $\{t_{i}\}_{i=1}^{n}$, this estimation is asymptotically unbiased, namely that \[\lim_{t\rightarrow\infty}E\{\widetilde{P}_{n}\}=P_{n}.\] While these assumptions are clearly unrealistic, we assume that as long as$\tau$ remains relatively small (i.e.~that the process is locally stationary within the kernel's time-window), this estimation is sufficiently accurate.

\subsection*{PID Controller}

A Proportional-Integral-Derivative (PID) controller was realized on the xPC target system. The input to the controller is the error signal,
\[e_{n}=R_{n}^{*}-R_{n}\]  where $R_{n}^{*}$ and $R_{n}$ are the desired and actual response to the $n^{th}$ stimulus, respectively. The output of the controller is generally composed of three components,

\[y_{n}=g_{P}e_{n}+g_{I}\sum_{i=0}^{n}e_{i}+g_{D}(e_{n}-e_{n-1})\]
 where $g_{P},g_{I}$ and $g_{D}$ are the proportional, integral and derivative gains, respectively (in the experiments presented in this work, $g_{P}$ was 25-30Hz, $g_{I}$ was 0.125-0.56Hz and $g_{D}$ was 0-125Hz). Finally, the instantaneous stimulation frequency (i.e. the reciprocal of the ISI) consisted of the controller's output and some 'baseline' frequency:

\[f_{n}=y_{n}+f_{baseline},\] where $f_{baseline}$ was 2Hz in latency experiments and 6.67Hz in probability experiments. In our system, the maximal stimulation rate was 40  Hz. In the case of failed spikes when latency was controlled, the controller maintained the same stimulation frequency that produced a spike at a previous stimulus.

\newpage

\section*{Results}

The experiments described in this work were conducted in networks of rat cortical neurons, plated \textit{in-vitro} on a substrate that caters to \textit{extracellular} recording and stimulation of electrical activity, at the single cell level. Evoked spiking activity of a given individual neuron in such  a network, reflects both the direct effect of external stimulation, as well as the effects of synaptic activations by other neurons \cite{marom2002development}.  We first examine the capacity of the \textit{Neuronal Response Clamp} technique to control the variability of neuronal responses of neurons that are \textit{isolated} from the network by  pharmacologically blocking all major types of synaptic inputs (APV [amino-5-phosphonovaleric acid; 20 $\mu$M], CNQX [6-cyano-7-nitroquinoxaline-2,3-dione; 10 $\mu$M], and Bicuculline [5 $\mu$M]). %Note that we do not address the issue of possible connection between neurons via gap-junctions, which are very rare in our system \cite{froes1999gap}.

Under complete blockade of synaptic transmission, all spontaneous activity dies out and the network becomes quiescent \cite{marom2002development}.  At this state, applying a single, short (sub-millisecond) pulse between two electrodes, generates a single action potential in a subset of neurons. The dynamics of the single neuron response may thus be interrogated using a long sequence of stimuli.  We refer to this stimulation scenario as an \emph{open loop} protocol, since the (predefined) stimulation sequence is not affected by the evoked neuronal responses. In the open loop protocol, the nature of the evoked responses leads to a distinction between two qualitatively different stimulation regimes: When stimulation rate is low (below 5-7 Hz), response probability is practically one, i.e.~the neuron responds almost to each and every stimulus, and the time-delay (latency) between the stimulus and the evoked spike is quite stable (Figure 1 middle panel).  At higher stimulation rates, both response probability and response latency gradually become unstable, exhibiting considerable, seemingly erratic fluctuations  (Figure 1 bottom).  Figure 2 depicts the dynamics of the two features (referred to collectively as the \emph{neural response}) at three stimulation rates (1, 5, and 20  Hz) over extended timescales.  Note the slow drift of response latency in the 5  Hz stimulation experiment, probably reflecting slow modulations of membrane excitability.  

Thus, two features of neural activity, \emph{response latency} and \emph{response probability}, emerge as natural candidates for stabilization using the Response Clamp technique. These two features are markers of the neuronal excitability status, reflecting the balance between exciting and restoring membrane conductances.  In order to clamp these two features, continuous, online estimation of their values is required.  Unlike response latency, which can be measured individually for each of the evoked action-potentials, response probability must be estimated based on recent spiking history (i.e.~using several responses). We calculate this latter measure by convolving the spiking history with an exponential kernel, which means that recent responses are given a greater weight (explained in Materials and Methods). %Note that multiplying the response probability (which is unit-less) by stimulation rate, yields the neuronal response rate (in units of $time^{-1}$).

The two response features mentioned above possess a useful quality, from the point of view of control:  they exhibit a simple, monotonic relationship with the instantaneous stimulation rate.  Increasing the rate (i.e.~by shortening the inter-stimulus-interval) leads to predictable outcomes, namely a decrease in spike probability and an increase in spike latency.  Therefore, negative feedback may be used in order to produce a desired response, either constant or time-varying. This is achieved in the following manner (Figure 3): A response feature to be clamped is chosen (either response probability or response latency).  Following each stimulus, a \emph{clamp error} is computed by comparing the evoked response with its desired value. This error signal is then fed into a PID controller which accordingly alters the stimulation rate in order to decrease the error towards zero. The PID controller consists of three components: the proportional component (P) supplies the necessary negative feedback; the integrated component (I) counteracts slow changes in the system's responsiveness; and the derivative component (D) contributes dissipation needed to damp oscillations.  For details on controller construction and implementation see Methods. 

%We now turn to the description of the stabilization of neural response using the above design. 
We first compared the neuronal response in open-loop, where stimulation rate is maintained constant, to the neuronal response under closed-loop conditions in synaptically isolated neurons for both response features, over a wide range of stimulation rates and response levels. In contrast to the open loop response, the response under closed-loop conditions quickly converges to the desired value (Figure 4); the clamp may be maintained for up to several hours (see below). Thus, Response Clamp counteracts the slow trends observed in the spike response latency, and restrains the huge fluctuations of the spike response probability.  The response clamp technique is reliable and robust: Figure 5 summarizes the results of 131 experimental blocks, showing that the standard deviation in response probability is markedly decreased under closed loop, compared to open loop stimulation regimes. The robustness of the technique is reflected in the fact that no fine-tuning of the PID gain parameters was required in order to achieve successful clamp. 

We were interested to determine whether it is the specific pattern of stimulation, evoked by the controller is sufficient, in and by itself to restrain response variability.  We therefore played-back a stimulation pattern produced by the controller under Response Clamp conditions, to stimulate the same neuron again in an open-loop manner.  Although the ``replayed'' input sequences were identical, stability quickly deteriorated and the response probability diverged from the desired value (Figure 6).  Thus, feedback is essential in order to achieve response stabilization: the exact temporal pattern of the stimulation series does not, in and by itself, cause stable neuronal response; rather, the instantaneous state of the neuron must be continuously monitored and taken into account while computing the control signal. Note that the resulting response \textit{rate} (i.e.~the product of stimulation rate and response probability), while not constant in either the open loop or under Response Clamp, is significantly more stable in the latter condition.  Also, note the duration of the experiment presented in Figure 6, extending over 3 hours. The control signal in this long experiment, as well as in similarly long experiments (results not shown), exhibits complex dynamics that expose the myriad underlying adaptive processes.

In all the above mentioned experiments the response was clamped to a \emph{constant} value. However, the neuronal response can also be clamped to a desired \emph{time-varying} pattern. To demonstrate this capacity, we performed an experiment where the desired response latency is gradually ramped up and down to three different values (Figure 7).  We observed that in the time-varying clamp, the rate at which the desired response may be changed is limited by the maximal and minimal stimulation rates; if the controller surpasses these values, its output saturates and the stimulation loop is effectively broken.  

In all the experiments described so far, neurons were pharmacologically isolated from the network input.  It was not at all obvious that the ability of the response clamp to restrain response variability would be conserved once ongoing input from the network is allowed.  We thus repeated the experiments described above \emph{without} blocking synaptic transmission, i.e.~when the neuron is embedded in an \textit{active} network.  We found that the Response Clamp technique was equally effective under these conditions (Figure 8).  Moreover, analyzing spikes simultaneously recorded from many different neurons in the network (beyond the controlled neuron), revealed that the responses of these other neurons fall into two broad categories:  Some neurons were influenced by the dynamic control over the target (controlled) neuron, whereas others seem to be completely unaffected.  These results suggest that the Response Clamp technique may be used as a tool to study \textit{single neuron} dynamics even when isolation of the neuron from its context is impossible or undesirable. %neuronal properties in the context of large-scale systems or to study how elementary dynamics affect the overall behavior of these large-scale systems.

\newpage

\section*{Discussion}

In this study we show that evoked neuronal spiking patterns may be controlled using a simple feedback system. The Response Clamp design was applied to control either the time delay between stimulus and an evoked spike, or the probability of evoked spikes. Control of these response variables was shown to be applicable both when the target neuron is synaptically isolated from the rest of the network, and on the background of ongoing synaptic input.

Historically, the concept of feedback control proved effective in the analyses of excitability.  Its use, however, was largely focused on the study of specific membrane conductances under voltage-clamp and patch-clamp conditions \cite{hodgkin1952measurement,neher1978extracellular}.  The approach was further advanced by the dynamic-clamp experimental design to unfold the impacts of specific conductances on the dynamics of the overall excitability \cite{sharp1993dynamic}.  In that context, the methodology presented here is a natural step up in the ladder of organization levels. It enables control of neuronal response patterns at the macroscopic level, without monitoring underlying microscopic variables.

One should bear in mind that in the original voltage-clamp studies of action potential generation the variable under control (namely, membrane potential) determines the reaction rates of \textit{all} the relevant processes, so that the closed loop behavior becomes linear and time invariant. In contrast, in the dynamic clamp method only a few components of the system are being controlled, with the hope of illuminating the role of these components in the overall behavior. In this context the method suggested here resembles the dynamic clamp in the sense that only the processes that depend on the responsiveness of the system are being clamped. Hopefully, this may aid in  the identification of these processes and the evaluation of their contribution to the response dynamics.

Two technical difficulties in the design of the Response Clamp must be noted. First, unlike the membrane potential which may be monitored continuously, the neuron must be externally perturbed in order to estimate its state (i.e.~responsiveness to stimulation). Therefore, the controller is always ``one step behind'' the system being controlled. Since these perturbations themselves affect the neuron's responsiveness, a classic ``observer effect'' arises.  This problem becomes more pronounced in the case of neural response probability, where the state is estimated using past spiking history.  Future improvements of the state estimation process might incorporate compensation for the effects of external stimulation.

A second limitation in implementing the Response Clamp technique relates to the response features to be controlled. The choice of response latency and response probability stems from the need for a predictable reaction to changes in stimulation rate. This does not necessarily require a monotonic relationship; a successful clamp may be realized as long as the short term outcomes of modifying the stimulation are predictable.

The Response Clamp technique offers the potential of facilitating full characterization of the input-output relationships of the neuron. This challenge is practically intractable in open loop due to the nonlinearity of the system and the cumulative effect of underlying processes spanning a wide range of  timescales. By manipulating well-defined features of the neuronal responsiveness, one may hope to control state-dependent dynamics, thus enabling the identification of the role of such processes in the overall behavior of the system.

%There are several straightforward directions for future extensions and applications of the Response Clamp technique offered here.  One option is to attempt controlling down-stream neurons, i.e.~neurons that are not directly affected by the external stimulation, yet receive synaptic inputs from neurons that are directly stimulated. This will extend the concept of response clamp to control of the dynamics of the total synaptic input to a given neuron. Another possibility is to study long-term dynamics of neural networks by controlling features at the population response level. % \cite{jimbo2000dynamics,marom2002development}. 
%Finally, it should be possible, in principle, to implement the Response Clamp technique on neuronal responses \emph{in-vivo}, both for understanding of, and intervention in sensory-motor loops in health and disease.

%Feedback paradigms have been applied in many studies at the behavioral level to manipulate neural responses (e.g. biofeedback). However, in such studies the control mechanism was supplied by the system itself, granting us with little information as to \emph{how} this mechanism operates; the Response Clamp may serve as a valuable tool in revealing how feedback control arises in the \emph{natural} system in order to stabilize desired responses during on-going activity.

%% Optional Appendix or Appendices
%% \appendix Appendix text...
%% or, for appendix with title, use square brackets:
%% \appendix[Appendix Title]
\newpage

\section*{Acknowledgments}
The authors thank Erez Braun, Naama Brenner, Danni Dagan, Steve Goldstein, Effraim Wallach and  Noam Ziv for their useful comments and suggestions.

\pagebreak
%\bibliographystyle{apacite}
%\bibliography{RC_Wallach_et_al}

\begin{thebibliography}{}

\bibitem[\protect\citeauthoryear{%
Adrian%
\ \BBA{} Zotterman%
}{%
Adrian%
\ \BBA{} Zotterman%
}{%
{\protect\APACyear{1926}}%
}]{%
Adrian:1926db}%
\APACinsertmetastar{%
Adrian:1926db}%
Adrian, E.~D.%
\BCBT{}\ \BBA{} Zotterman, Y.%
%
\newblock{}\BBOP{}1926\BBCP{}.
\newblock{}\BBOQ{}The impulses produced by sensory nerve endings: Part 3.
  impulses set up by touch and pressure.\BBCQ{}
\newblock{}\Bem{J Physiol}, \Bem{61}(4), 465--483.

\bibitem[\protect\citeauthoryear{%
Arieli%
, Sterkin%
, Grinvald%
\BCBL{}\ \BBA{} Aertsen%
}{%
Arieli%
\ \protect\BOthers{.}}{%
{\protect\APACyear{1996}}%
}]{%
arieli1996dynamics}%
\APACinsertmetastar{%
arieli1996dynamics}%
Arieli, A.%
, Sterkin, A.%
, Grinvald, A.%
\BCBL{}\ \BBA{} Aertsen, A.%
%
\newblock{}\BBOP{}1996\BBCP{}.
\newblock{}\BBOQ{}{Dynamics of ongoing activity: explanation of the large
  variability in evoked cortical responses}.\BBCQ{}
\newblock{}\Bem{Science}, \Bem{273}(5283), 1868.

\bibitem[\protect\citeauthoryear{%
Arsiero%
, Luscher%
, Lundstrom%
\BCBL{}\ \BBA{} Giugliano%
}{%
Arsiero%
\ \protect\BOthers{.}}{%
{\protect\APACyear{2007}}%
}]{%
arsiero2007impact}%
\APACinsertmetastar{%
arsiero2007impact}%
Arsiero, M.%
, Luscher, H.%
, Lundstrom, B.%
\BCBL{}\ \BBA{} Giugliano, M.%
%
\newblock{}\BBOP{}2007\BBCP{}.
\newblock{}\BBOQ{}{The impact of input fluctuations on the frequency-current
  relationships of layer 5 pyramidal neurons in the rat medial prefrontal
  cortex}.\BBCQ{}
\newblock{}\Bem{Journal of Neuroscience}, \Bem{27}(12), 3274.

\bibitem[\protect\citeauthoryear{%
Hodgkin%
, Huxley%
\BCBL{}\ \BBA{} Katz%
}{%
Hodgkin%
\ \protect\BOthers{.}}{%
{\protect\APACyear{1952}}%
}]{%
hodgkin1952measurement}%
\APACinsertmetastar{%
hodgkin1952measurement}%
Hodgkin, A.%
, Huxley, A.%
\BCBL{}\ \BBA{} Katz, B.%
%
\newblock{}\BBOP{}1952\BBCP{}.
\newblock{}\BBOQ{}{Measurement of current-voltage relations in the membrane of
  the giant axon of Loligo}.\BBCQ{}
\newblock{}\Bem{The Journal of physiology}, \Bem{116}(4), 424.

\bibitem[\protect\citeauthoryear{%
Levine%
\ \protect\BOthers{.}}{%
Levine%
\ \protect\BOthers{.}}{%
{\protect\APACyear{1996}}%
}]{%
levine1996control}%
\APACinsertmetastar{%
levine1996control}%
Levine, W.%
\BCBT{}\ \BOthersPeriod{.}%
\newblock{}\BBOP{}1996\BBCP{}.
\newblock{}\Bem{{The Control Handbook}}.
\newblock{}CRC press Boca Raton, FL.

\bibitem[\protect\citeauthoryear{%
Mainen%
\ \BBA{} Sejnowski%
}{%
Mainen%
\ \BBA{} Sejnowski%
}{%
{\protect\APACyear{1995}}%
}]{%
mainen1995reliability}%
\APACinsertmetastar{%
mainen1995reliability}%
Mainen, Z.%
\BCBT{}\ \BBA{} Sejnowski, T.%
%
\newblock{}\BBOP{}1995\BBCP{}.
\newblock{}\BBOQ{}{Reliability of spike timing in neocortical neurons}.\BBCQ{}
\newblock{}\Bem{Science}, \Bem{268}(5216), 1503.

\bibitem[\protect\citeauthoryear{%
Marom%
}{%
Marom%
}{%
{\protect\APACyear{2010}}%
}]{%
Marom201016}%
\APACinsertmetastar{%
Marom201016}%
Marom, S.%
%
\newblock{}\BBOP{}2010\BBCP{}.
\newblock{}\BBOQ{}Neural timescales or lack thereof.\BBCQ{}
\newblock{}\Bem{Progress in Neurobiology}, \Bem{90}(1), 16 - 28.

\bibitem[\protect\citeauthoryear{%
Marom%
\ \BBA{} Shahaf%
}{%
Marom%
\ \BBA{} Shahaf%
}{%
{\protect\APACyear{2002}}%
}]{%
marom2002development}%
\APACinsertmetastar{%
marom2002development}%
Marom, S.%
\BCBT{}\ \BBA{} Shahaf, G.%
%
\newblock{}\BBOP{}2002\BBCP{}.
\newblock{}\BBOQ{}{Development, learning and memory in large random networks of
  cortical neurons: lessons beyond anatomy}.\BBCQ{}
\newblock{}\Bem{Quarterly Reviews of Biophysics}, \Bem{35}(01), 63--87.

\bibitem[\protect\citeauthoryear{%
Neher%
, Sakmann%
\BCBL{}\ \BBA{} Steinbach%
}{%
Neher%
\ \protect\BOthers{.}}{%
{\protect\APACyear{1978}}%
}]{%
neher1978extracellular}%
\APACinsertmetastar{%
neher1978extracellular}%
Neher, E.%
, Sakmann, B.%
\BCBL{}\ \BBA{} Steinbach, J.%
%
\newblock{}\BBOP{}1978\BBCP{}.
\newblock{}\BBOQ{}{The extracellular patch clamp: a method for resolving
  currents through individual open channels in biological membranes}.\BBCQ{}
\newblock{}\Bem{Pflugers Archiv European Journal of Physiology}, \Bem{375}(2),
  219--228.

\bibitem[\protect\citeauthoryear{%
Reich%
, Victor%
, Knight%
, Ozaki%
\BCBL{}\ \BBA{} Kaplan%
}{%
Reich%
\ \protect\BOthers{.}}{%
{\protect\APACyear{1997}}%
}]{%
reich1997response}%
\APACinsertmetastar{%
reich1997response}%
Reich, D.%
, Victor, J.%
, Knight, B.%
, Ozaki, T.%
\BCBL{}\ \BBA{} Kaplan, E.%
%
\newblock{}\BBOP{}1997\BBCP{}.
\newblock{}\BBOQ{}{Response variability and timing precision of neuronal spike
  trains in vivo}.\BBCQ{}
\newblock{}\Bem{Journal of neurophysiology}, \Bem{77}(5), 2836.

\bibitem[\protect\citeauthoryear{%
Sharp%
, O'Neil%
, Abbott%
\BCBL{}\ \BBA{} Marder%
}{%
Sharp%
\ \protect\BOthers{.}}{%
{\protect\APACyear{1993}}%
}]{%
sharp1993dynamic}%
\APACinsertmetastar{%
sharp1993dynamic}%
Sharp, A.%
, O'Neil, M.%
, Abbott, L.%
\BCBL{}\ \BBA{} Marder, E.%
%
\newblock{}\BBOP{}1993\BBCP{}.
\newblock{}\BBOQ{}{Dynamic clamp: computer-generated conductances in real
  neurons}.\BBCQ{}
\newblock{}\Bem{Journal of neurophysiology}, \Bem{69}(3), 992.

\bibitem[\protect\citeauthoryear{%
Soteropoulos%
\ \BBA{} Baker%
}{%
Soteropoulos%
\ \BBA{} Baker%
}{%
{\protect\APACyear{2009}}%
}]{%
soteropoulos2009quantifying}%
\APACinsertmetastar{%
soteropoulos2009quantifying}%
Soteropoulos, D.%
\BCBT{}\ \BBA{} Baker, S.%
%
\newblock{}\BBOP{}2009\BBCP{}.
\newblock{}\BBOQ{}{Quantifying Neural Coding of Event Timing}.\BBCQ{}
\newblock{}\Bem{Journal of Neurophysiology}, \Bem{101}(1), 402.

\bibitem[\protect\citeauthoryear{%
Stein%
}{%
Stein%
}{%
{\protect\APACyear{1965}}%
}]{%
stein1965theoretical}%
\APACinsertmetastar{%
stein1965theoretical}%
Stein, R.%
%
\newblock{}\BBOP{}1965\BBCP{}.
\newblock{}\BBOQ{}{A theoretical analysis of neuronal variability}.\BBCQ{}
\newblock{}\Bem{Biophysical Journal}, \Bem{5}(2), 173--194.

\end{thebibliography}

%%%%%%%%%%%%%%%%%%%%%%%%%%%%%%%%%%%%%%%%%%%%%%%%%%%%%%%%%%%%%%%%
\newpage

\section*{Figures}

\begin{figure*}
	\centering
	\includegraphics[width=5 in]{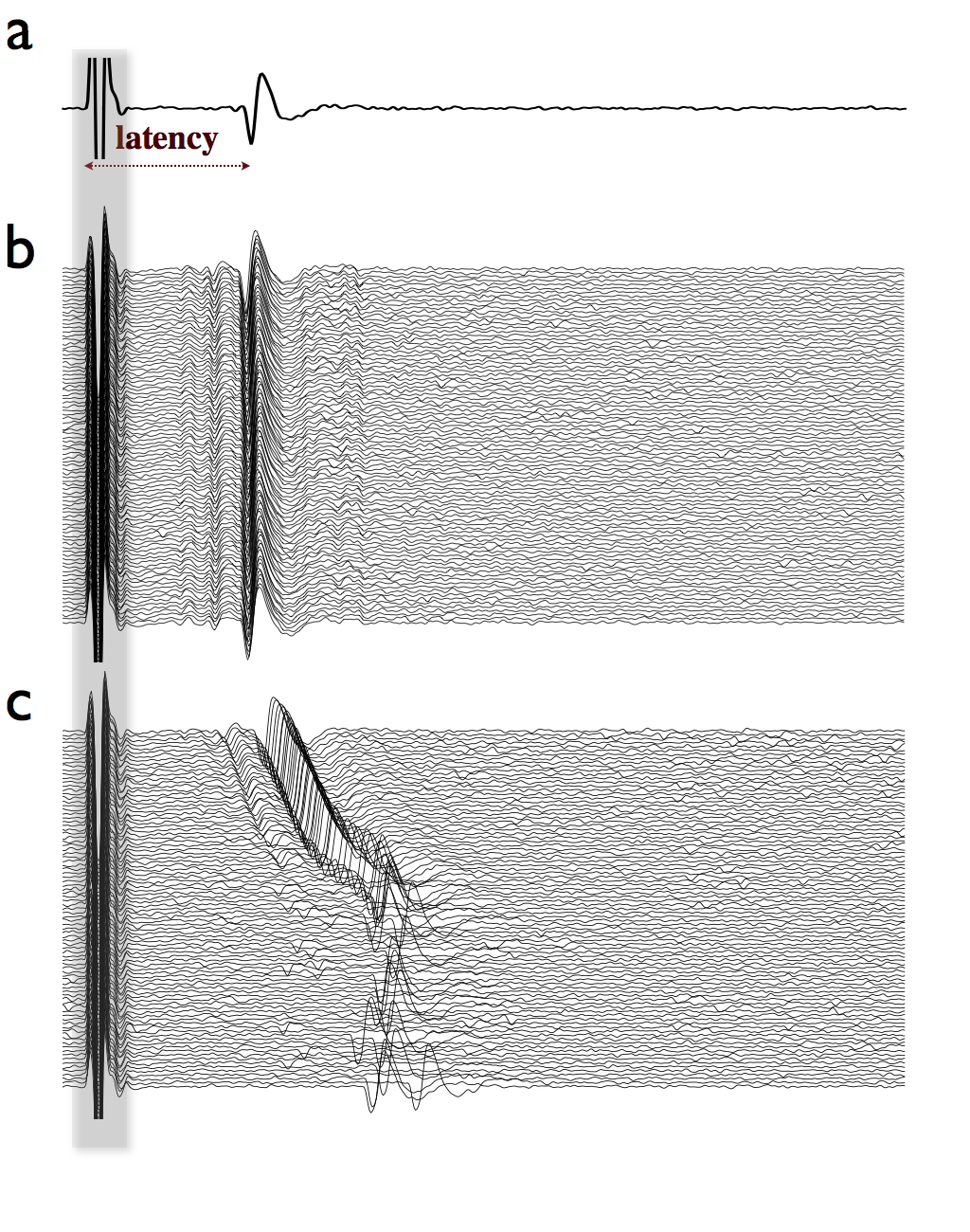} 
 	  \caption{Evoked responses of an isolated neuron \textit{in-vitro} to periodic stimulation in open loop. The top panel (\textbf{a}) shows a single extracellular voltage trace recorded for 20 ms, beginning at the onset of stimulation. The stimulus artifact lasts for 1-2 ms (shaded area). Response latency is the time elapsed from the onset of stimulation to the detected peak of the action potential; in our system, response latency is in the range of 3-10  ms. (\textbf{b}) Traces (every 10th, shown top to bottom) in response to a low (1 Hz) stimulation rate. Spiking under this condition is highly regular and the latency is  constant. (\textbf{c}) Traces (every 10th, shown top to bottom) of the same neuron when stimulation rate is high (20 Hz). Both response probability and latency rapidly become irregular under this condition.}
	\label{Figure 1}
\end{figure*}

\begin{figure*}
	\centering
	\includegraphics[width=5 in]{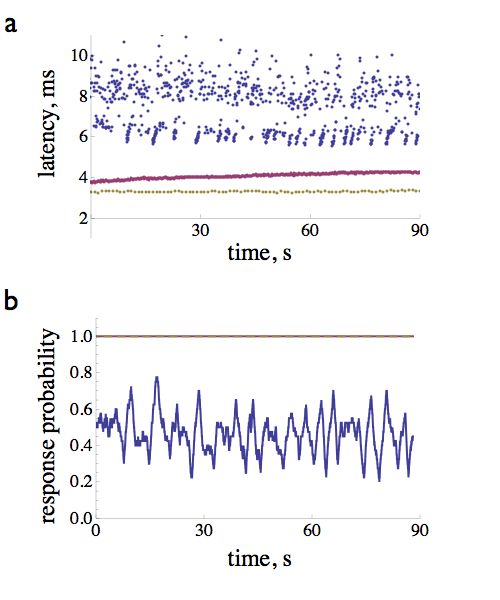} 
 	  \caption{Response latency (\textbf{a}) and response probability (\textbf{b}) for three open loop stimulation rates. At extremely low stimulation rate (1 Hz, depicted yellow), response is highly reliable (i.e.~response probability is one) and response latency is practically constant. At higher stimulation rate (5 Hz, depicted purple), response probability is also one, but slow trends in the response latency appear, probably reflecting slow processes of inactivation. At high stimulation rate (20 Hz, depicted blue) both response probability and response latency are highly unstable.  The response probability was computed using 2 s bins.}
	\label{Figure 2}
\end{figure*}

\begin{figure*}
	\centering
	\includegraphics[width=5 in]{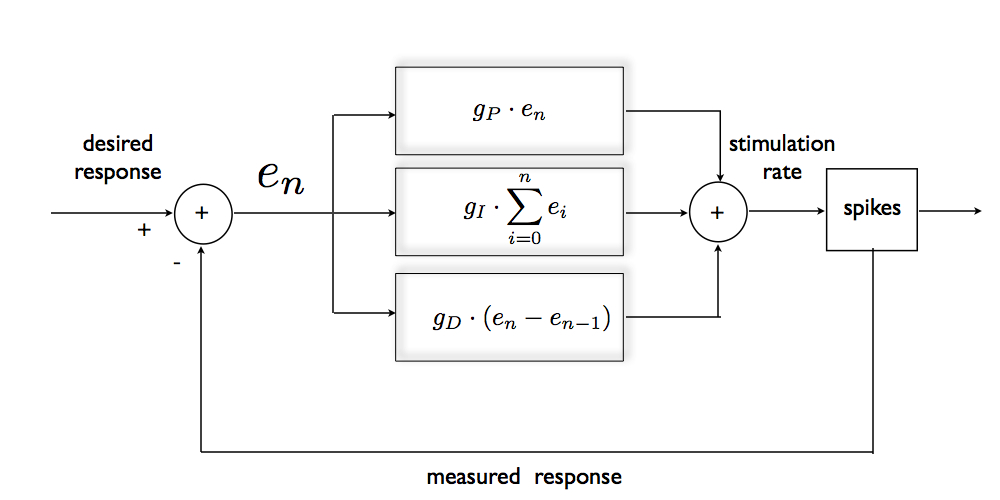} 
 	  \caption{General scheme of a PID controller, designed for clamping neuronal responses (stimulus-spike time delay, or response probability). The error signal is calculated (feedback), and subjected to three different transformations that additively dictate the nature of stimulation needed for clamping the response.  This is standard control algorithm was implemented within a \textit{Simulink} environment.}
	\label{Figure 3}
\end{figure*}

\begin{figure*}
	\centering
	\includegraphics[width=5 in]{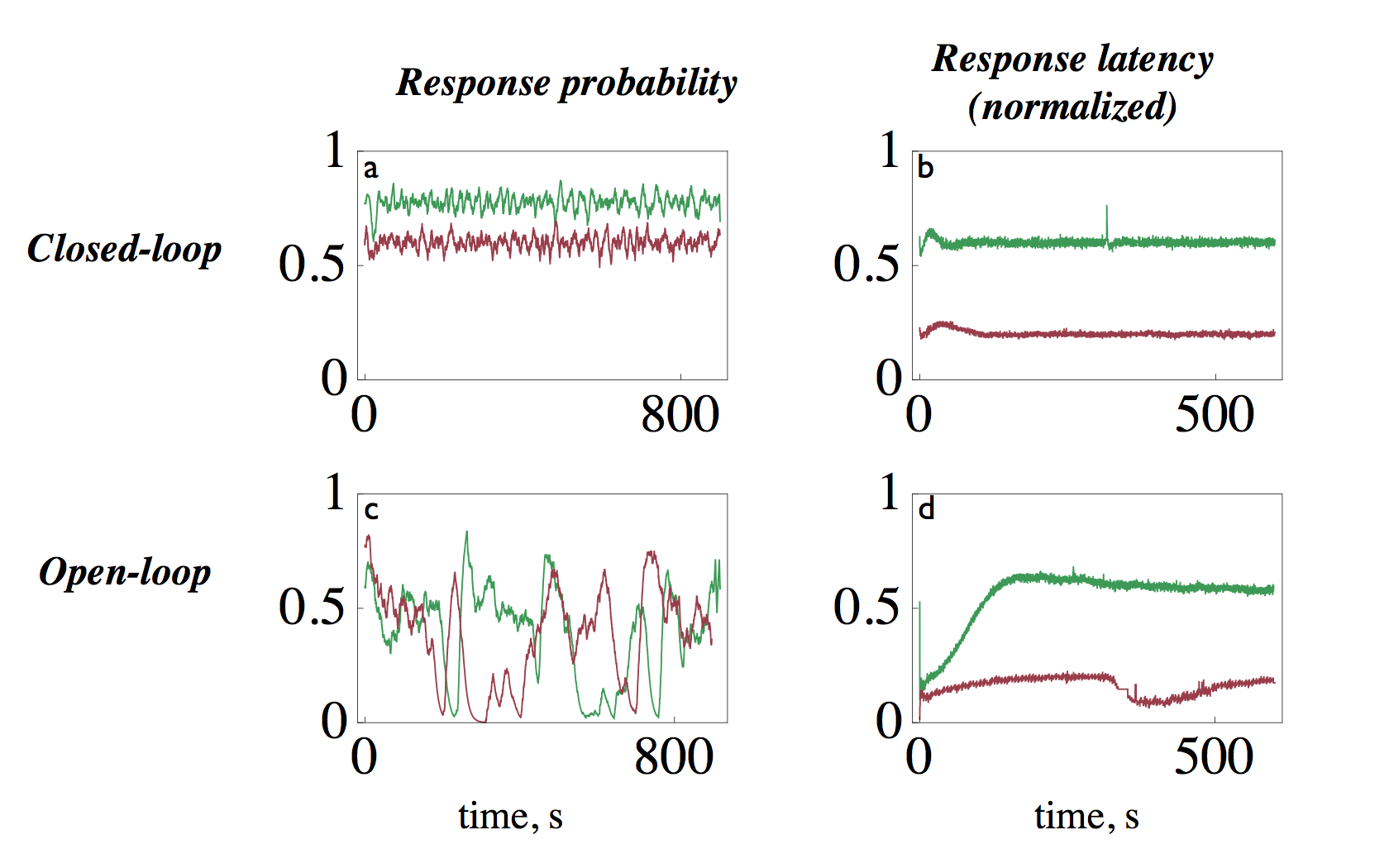} 
 	  \caption{
	Demonstration of the neuronal Response Clamp.  In each of the panels, two experiments are shown (depicted by two different colors).  Top row shows examples of controlling response probability (\textbf{a}) and stimulus-spike time delay (\textbf{b}).  Two clamped values are demonstrated in each of these panels.  Note the stability of the clamped response.  In contrast, when the average stimulation rates used for the clamping procedure of the two top panels are applied under open loop conditions (\textbf{c} and \textbf{d}), the response develops marked fluctuations.  This is  especially apparent when response probability (\textbf{c}) is considered. 
	}
	\label{Figure 4}
\end{figure*}

\begin{figure*}
	\centering
	\includegraphics[width=5 in]{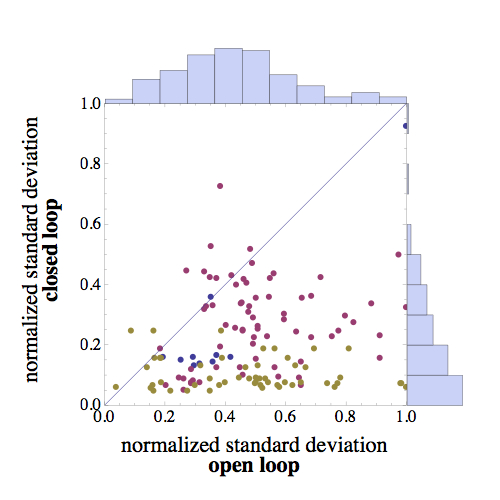} 
 	  \caption{
The Response Clamp consistently decreases the variability of response probability. Neuronal response probability was clamped to different values ranging from 1 to 0.25, for 10 to 15 minutes.  The response's standard deviation during this period is compared with that of the same neuron, when the mean rate used during the clamp period is re-applied in an open loop design (see the examples depicted in Figure 4). Data from 131 such experimental sessions conducted on 3 different neurons (each in a different culture, depicted by different colors) are presented. The data are normalized to the maximal standard deviation acquired for each neuron.  Histograms of the data are aligned to the relevant axes.}
	\label{Figure 5}
\end{figure*}

\begin{figure*}
	\centering
	\includegraphics[width=5 in]{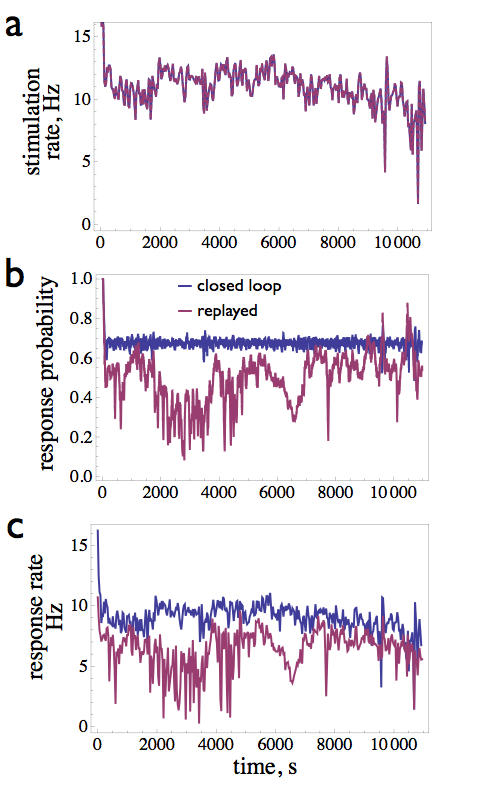} 
 	  \caption{
	Demonstration of the importance of feedback in the stabilization of neuronal response probability. Stimulation rate (\textbf{a}), response probability (\textbf{b}) and response rate (\textbf{c}) at two stimulation scenarios are depicted. First, the neuron's response probability was clamped to a constant value (0.67) for three hours (blue). Then, the stimulation pattern generated by the controller was replayed in open loop (purple). Stability of both response probability and response rate is obtained only in the presence of feedback.
	}
	\label{Figure 6}
\end{figure*}

\begin{figure*}
	\centering
	\includegraphics[width=5 in]{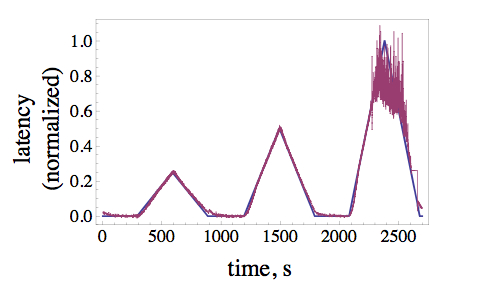} 
 	  \caption{
Demonstration of time varying Response Clamp. Response latency was clamped, while the desired value (depicted blue) was gradually increased and decreased alternatingly. Latency is normalized so that the baseline level (i.e.~the latency at an extremely low stimulation rate, in the example presented here it is 6  ms) is 0 and maximal latency detected (10  ms in this example) is set to 1. Note that at extremly high latency values the behavior becomes irregular and the clamp is effectively lost.
}
	\label{Figure 7}
\end{figure*}

\begin{figure*}
	\centerline{\includegraphics[width=5 in]{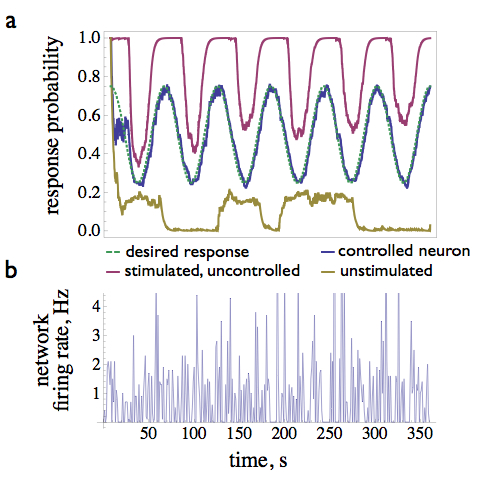} }
 	  \caption{
	Time varying Response Clamp of a neuron with background synaptic input. (\textbf{a}) A neuron that responds (blue) directly to external stimulation was controlled to follow a sine-wave response probability pattern (12 minutes period, range between 0.75 and 0.25, depicted light green). The figure shows the response probability of another, directly stimulated yet uncontrolled neuron (purple). The response of such neurons, while not follwing the desired response closely like the controlled neuron does, is modulated by the clamp. The figure also presents the response probability of yet another neuron, which is not affected directly by the stimulation, but is activated by the synaptic inputs it recieves from the network (yellow). The activity of such neurons seems to be unaffected by the clamp. This is also apparent by looking at the spike rate histogram of 20 other neurons in the network (\textbf{b}, 10 s bins), demonstrating the presence of uninterrupted on-going background activity in the network. 
	}
	\label{Figure 8}
\end{figure*}

%%%%%%%%%

\end{document}